\documentstyle[12pt]{article}
\input{epsf}
\def\m{\mu}
\def\n{\nu}
\catcode`\@=11
\newcount\hour
\newcount\minute
\newtoks\amorpm
\hour=\time\divide\hour by60
\minute=\time{\multiply\hour by60 \global\advance\minute by-\hour}
\edef\standardtime{{\ifnum\hour<12 \global\amorpm={am}%
        \else\global\amorpm={pm}\advance\hour by-12 \fi
        \ifnum\hour=0 \hour=12 \fi
        \number\hour:\ifnum\minute<10 0\fi\number\minute\the\amorpm}}
\edef\militarytime{\number\hour:\ifnum\minute<10 0\fi\number\minute}
\def\draftlabel#1{{\@bsphack\if@filesw {\let\thepage\relax
   \xdef\@gtempa{\write\@auxout{\string
      \newlabel{#1}{{\@currentlabel}{\thepage}}}}}\@gtempa
   \if@nobreak \ifvmode\nobreak\fi\fi\fi\@esphack}
        \gdef\@eqnlabel{#1}}
\def\@eqnlabel{}
\def\@vacuum{}
\def\draftmarginnote#1{\marginpar{\raggedright\scriptsize\tt#1}}
\def\draft{\oddsidemargin -.2truein
        \def\@oddfoot{\sl preliminary draft \hfil
        \rm\thepage\hfil\sl\today\quad\militarytime}
        \let\@evenfoot\@oddfoot \overfullrule 3pt
        \let\label=\draftlabel
        \let\marginnote=\draftmarginnote
   \def\@eqnnum{(\theequation)\rlap{\kern\marginparsep\tt\@eqnlabel}%
\global\let\@eqnlabel\@vacuum}  }
\relax
\def\ads{AdS}

\renewcommand{\theequation}{\arabic{section}.\arabic{equation}}

\def\be{\begin{equation}}
\def\ee{\end{equation}}
\def\bs{\begin{subequations}}
\def\es{\end{subequations}}

\newcommand{\prl}{Phys.\ Rev.\ Lett.~}
\newcommand{\pr}{Phys.\ Rev.\ D~}

\newcommand{\np}{Nucl.\ Phys.\ B~}

\begin{document}

\begin{titlepage}
\begin{flushright}
LPTHE--ORSAY 98/54\\
hep-th/9906048
\end{flushright}
\vspace{.0in}

\begin{center}
{\Large\bf Thermodynamics of D-brane Probes\footnote{\small Research
supported in part by
the National Science Foundation under grant
PHY--96--02074, and in part by the EEC
under the TMR contract
ERBFMRX-CT96-0090.}}
\vskip 1cm

{\large E. Kiritsis$^{\,a}$
and T.R. Taylor$^{\,b}$}
\vspace{.5in}

$^a$Physics Department, University of Crete\\ P.O. Box 2208,  71003 Heraklion, GREECE\\
\vskip .7 cm
$^b$Department of Physics, Northeastern
University\\ Boston, MA 02115, U.S.A.

\end{center}

\vskip .15in

\begin{center} {\bf ABSTRACT }
 \end{center}
\begin{quotation}\noindent

We discuss the dynamics and thermodynamics of particle and D-brane
probes
moving in non-extremal black hole/brane backgrounds. When a probe falls
from asymptotic infinity to the  horizon, it transforms its potential
energy into heat, $TdS$, which is absorbed by the black hole in a way
consistent with the first law of thermodynamics.
We show that the same remains true in the near-horizon limit, for BPS probes only,
with the BPS probe moving from AdS
infinity to the horizon.
This is a quantitative indication that the brane-probe reaching the
horizon corresponds to thermalization in gauge theory.
It is shown that this relation provides a way to reliably compute the entropy 
away from
the extremal limit (towards the Schwarzschild limit).
\end{quotation}

\vskip 1cm

\centerline{\tt Based on talks presented at the midterm meeting of the TMR network}
\centerline{\tt "Physics beyond
the standard model," held in Trieste in March 1999,}
\centerline{\tt and at the 1998 Corfu Summer Institute on 
Elementary Particle Physics.}
\centerline{\tt To appear in the proceedings.}

\vskip 2cm
\begin{flushleft}
June 1999\\
\end{flushleft}

\end{titlepage}
\vfill
\eject

\def\baselinestretch{1.2}
\baselineskip 14 pt
\noindent

\setcounter{section}{0}
\section{Introduction and conclusions}

The black D-brane solutions of type II supergravity \cite{hs} and their
near-horizon geometry \cite{gt} are the central elements of
CFT/Anti-deSitter (AdS) correspondence \cite{mald,gkp,w1}.
In particular, the (3+1)-dimensional world-volume of
$N$ coinciding, extremal D3-branes is the arena of ${\cal N}{=}4$
supersymmetric $SU(N)$ Yang-Mills (SYM) theory which in the large $N$
limit,
according to the Maldacena conjecture \cite{mald},
is linked  to type IIB superstrings propagating on (the
near-horizon) $\ads_5\times S^5$ background geometry.
Recently, there has been an interesting proposal \cite{w2} linking the
thermodynamics  of large $N$, ${\cal N}{=}4$ supersymmetric
$SU(N)$ Yang-Mills  theory with the thermodynamics of Schwarzschild
black holes embedded in the AdS space \cite{hp}.
The classical geometry of black holes with Hawking temperature $T$
encodes the magnetic confinement, mass gap and other qualitative
features of large $N$ gauge theory heated
up to the same temperature. At the computational level,
the quantity that has been discussed to the largest
extent \cite{gkpeet,w2,gkt} is the Bekenstein-Hawking entropy
which, in the near-horizon limit,
should be related to the entropy of Yang-Mills gas at
$N\rightarrow\infty$ and large 't Hooft coupling $g^2_{\rm YM}N$.

The SYM/AdS correspondence and its thermal black hole
generalizations emerge in a particular limit of $N$ D-branes
coinciding at one point in the transverse space; this
corresponds to a conformal point in the SYM moduli space,
with zero vacuum expectation values (vevs) of all scalar fields.
Before taking the near-horizon limit one could also consider
some other configurations, obtained for instance by placing a number of
D-branes in the bulk of AdS;
this corresponds to switching on some scalar vevs
on the Higgs branch of the gauge theory. In this work,
we consider the case of a single D-brane probe in the background
of a {\em near}-extremal black hole with a large number of coinciding
D-branes.
We consider a probe moving from asymptotic infinity towards the
black hole horizon. As the probe moves through the horizon,
the black hole receives the quantity of heat that is determined by
the first law of thermodynamics. The corresponding change in black hole
entropy
is consistent with thermodynamical identities.
Hence from the $thermodynamical$ point of view,
the D-brane gas is physically located at the black hole horizon.

We can then consider a similar process in the near-horizon limit
description.
That is, a probe brane is falling from the boundary of the appropriate
space (AdS$_5$ for D3-branes) rather than from spatial infinity,
to the horizon.
We find again, for BPS probes,  that the potential energy released is equal to the heat,
$TdS$, absorbed by the black hole.
Thus, this process can be used to calculate the entropy of gauge theory
in an alternative way by integrating back the potential energy.
Moreover, this process has now an interpretation in terms of the
spontaneously
broken $SU(N+1)\to SU(N)\times U(1)$ gauge theory.
At $T=0$, any Higgs expectation value is stable, since it is a modulus
protected by supersymmetry.
Once $T>0$, supersymmetry is broken, and the Higgs acquires a potential
with an absolute minimum at zero expectation value.
Starting with the theory at the top of the potential (very large
expectation
value) the Higgs will start rolling  down.
The Higgs field reaching an expectation value of the order of
the electric mass corresponds in supergravity to
the probe reaching the horizon.
At this point, the energy stored in the Higgs field is thermalized and equal
to the
overall heat received by the thermal Yang-Mills system.
This makes more precise a similar picture described in \cite{bdhm}.
{}From the supergravity point of view the brane continues to move,
crossing the
horizon.
It is not obvious what this motion corresponds to in gauge theory (but see
\cite{bdhm} for a proposal).

We consider also the effects of higher order $\alpha'$ corrections to the probe action.
Such corrections have been partially computed for the appropriate backgrounds
directly
in the near-horizon limit \cite{bbg}.
We argue that that although there are $\alpha'$ corrections to the
D-brane
world-volume action, they do not influence the calculation of the heat
in our argument.

In all of the examples analyzed here, the probe method can  be viewed
as an independent
method of ``measuring the entropy" by integrating the heat, $TdS$.
In particular, it provides a first order partial differential equation
for the entropy in terms of mass $M$ and charge $Q$.
$\alpha'$-corrections 
do not modify the leading equation.
The only source of stringy corrections to the entropy equation comes 
from bulk $\alpha'$-corrections to the RR gauge potential at the horizon.
This equation is important since it can be used in conjunction with
entropy calculations, done near the limit where supersymmetry is
restored
(and thus the calculation is reliable) and then extrapolated to the
Schwarzschild
region where supersymmetry is completely broken and
calculations are difficult to control.

This talk is based on results obtained in collaboration with Constantin Bachas
\cite{bktt}. We will first describe in detail the case of 
Reissner-Nordstr\"om black holes 
in four dimensions and their near-horizon limit.
Then we will describe arbitrary black Dp-branes and finally focus on the 
near-horizon limit of D3-branes.

\section{Reissner-Nordstr\"om black holes}

In order to see   how probes can be used to study  black hole
thermodynamics, it is instructive to consider first the
case of a  point  particle
propagating in the  background geometry of a charged
black hole. The Reissner-Nordstr\"om (RN) metric for a charged black hole
with  ADM mass $M$ and electric
charge $Q$  reads
\be ds^2=-\left(1-{2l_p^2M\over r}+{l^2_pQ^2\over
r^2}\right)dt^2+\left(1-{2l_p^2M\over r}+{l_p^2Q^2\over
r^2}\right)^{-1}dr^2+r^2d\Omega_2^2  \label{rn}
\ee
where $G_N=l_p^2$ is Newton's constant.
There is also
a static electromagnetic potential, which can be obtained from Gauss'
law, $A_0={Q\over r}$.
The outer and inner  horizons are located at $r_+$ and $r_-$ with
$r_{\pm}=l_p^2\left(M\pm \sqrt{M^2-Q^2/l_p^{2}}\right)$.
The standard expression for the Bekenstein-Hawking entropy,
\be
S(M,Q) = {{\rm Area}  \over 4G_N} = {\pi r_+^2\over l_p^2} \ ,
\ee
can be thought of as
  the `equation of state' in the microcanonical ensemble. From
it we can obtain  the  temperature and  `chemical
potential' of the black hole
\be
{1\over T} \equiv  {\partial S\over \partial M}\Biggl\vert_Q =
 { 4\pi r_+^2\over {r_+-r_-}}
\;,\;\;\;
\mu \equiv  -T  {\partial S\over \partial Q}\Biggr\vert_M =   {Q\over
r_+}\
{}.
\label{par}
\ee
The free energy and grand canonical potential can also be obtained by
the standard thermodynamic expressions,   $F \equiv  M -
TS$ and $A \equiv  M - TS - \mu  Q$.

  Consider next the process in which a probe particle with mass
$m$ and electric charge $q$ falls inside the black hole. The black
hole plus particle form an isolated system,  with total mass $M+m$ and
total charge $Q+q$, which will eventually reach thermal equilibrium.
The entropy
will therefore change  by an amount
\be
 \delta S = {\partial S\over \partial M}\Biggl\vert_Q m + {\partial
 S\over \partial  Q}\Biggl\vert_M q \ . \label{hea}
\ee
Using the explicit form of the chemical potential,  derived  from
the `equation of state', we find
\be
T \delta S = m - {Qq \over r_+}\ . \label{heat}
\ee
This equation admits a simple interpretation as the heat released by
the
probe particle while falling inside the  black hole.
The action for  such a particle indeed reads
\be
\Gamma=  m\int d\tau \sqrt{-G_{\mu\nu}\dot x^{\mu}\dot x^{\nu}
}+q\int
d\tau
A_{\mu}\dot x^{\mu}\ .
\ee
In the static gauge,
 $t=\tau$, it takes the form
\be
\Gamma=\int dt~ V(r)+{\rm velocity~~terms}
\label{pact}\ee
with  $V$  the static potential,
\be
V(r)=m\sqrt{\left(1-{2l_p^2M\over r}+{l_p^2Q^2\over
r^2}\right)}+{qQ\over r} \ .
\label{potential}
\ee
Notice that the  potential includes the self-energy $m$ of the probe,
and is constant in the extremal limit for both source ($M=Q/l_p$)  and
probe
($m=q/l_p$), consistently with the absence of a  static force in this
case.
Now as the particle moves from spatial infinity to the outer horizon,
the difference in  potential energy, $\delta V = V(\infty) - V(r_+)$,
is converted to kinetic
energy  and then  eventually dissipated  as heat. This is precisely the
content of eq.
(\ref{heat}).

  The argument can also  be run backwards. Starting from the static
  potential  eq. (\ref{potential}), and
  assuming that the potential energy of the probe  is converted to heat
  at the outer horizon,
 leads to eq. (\ref{heat}).
Comparing with eq. (\ref{hea})  then gives
\be
\mu = A_0(r_+) = {Q\over r_+}\ \ \ {\Longleftrightarrow}\ \ \
- {\partial
 S\over \partial  Q}\Biggl\vert_M =  A_0(r_+) \;
{\partial S\over \partial M}\Biggl\vert_Q\ . \label{diff}
\ee
This partial differential equation can be integrated for the equation
of state $S(M,Q)$, provided we know already the answer  on  some
(initial)  curve in  the $(M,Q)$ plane, such as for
Schwarzschild black holes $Q=0$ or extremal black holes, $M=Q/l_p$. Notice that even
though the thermodynamic properties of a Schwarzschild black hole are
in an essential way quantum ($\hbar$  enters in the expressions for
both temperature
and entropy), the extension to charged black holes follows from
the  simple classical argument  outlined here.
Such a differential equation may turn out to be practically important.
Several thermodynamic properties seem to be more easily computable
on or close to the extremal limit. There, supersymmetry is of help as
evidenced by recent D-brane/black hole calculations \cite{bhc}.
Being able to calculate at the extremal boundary of the $(M,Q)$ plane, one
can use (\ref{diff}) to extrapolate the calculation to the whole plane,
most importantly in the region $Q=0$ where supersymmetry is completely
broken.
It should be noted though that imposing a boundary condition at the
extremal boundary $M=Q$ maybe problematic since in most cases the
partial derivatives diverge there ($T=0$). This is the case here as well as
for
the Dp-branes with $p<5$.
We could however impose boundary conditions on a line just outside the
extremal boundary (at near-extremality) where computations are still
reliable.
This point is conceptually important and needs further investigation.

 The reader may of course  object  that these considerations
depend on our choice of a
minimal probe action.  Quantum gravity effects or stringy effects
(controlled respectively by $l_p$ and $l_s$)
  can give rise to
curvature terms and/or   non-minimal electromagnetic couplings,
which would  modify  the  potential (\ref{potential}). Nevertheless,
 {\it assuming thermodynamic equilibrium},  the relation
\be
{ T\delta S = V(\infty) - V(r_+) = m - q A_0(r_+) \  \label{genrl}
}
\ee
continues,  as we will now argue, to be  valid.
 This is of course
consistent with the fact that for a neutral particle $T\delta S = m$
is simply the first law of thermodynamics.

To see why the potential difference is always given by eq.(\ref{genrl}), 
consider possible corrections to the  action $\Gamma$ of the
particle.
These must be of the form
\be{
\delta\Gamma = \int ds \; f(R, F, {\dot x})\ , \label{modif}
}
\ee
 with $ds \equiv d\tau \sqrt{{\dot x}^\mu
{\dot x}_\mu}$
 the invariant proper time element,
and $f$ a  scalar made out of the electromagnetic field
strength, the Riemann tensor,  the probe velocity,  and covariant
derivatives  thereof. Note that corrections of the form $\int dx^\mu
A_\mu  {\tilde f}$ are not allowed, because gauge invariance forces the
scalar function $\tilde f$  to be a constant.\footnote{This does not
imply that the
electromagnetic coupling must be minimal, since (\ref{modif}) may
depend non-trivially  on the field strength.} Now in
 the quasi-static limit  $x^\mu = (\tau,
0\cdots 0)$, the invariant element $ds$   tends to $dt$ at spatial
infinity, and
vanishes  at the horizon,  when expressed in terms of the asymptotic
time. The scalar
$f$ on the other hand must  vanish  at spatial infinity where
both $F$ and $R$ go to zero. Thus, {\it provided   $f$ stays smooth
at
the event horizon},  these higher-order terms will  not contribute
to the static potential either at $r=r_+$
or at  $r=\infty$, as advertised.

To show that $f$ cannot indeed diverge at the horizon, note first that
this is
automatic  for  a scalar  function of the background fields,
since the horizon singularity is a  coordinate artifact.
Suppose next  that  $f$ is the pull back on the world-line of a
space-time tensor
\be
f = T_{\mu_1\cdots \mu_{n}}\;  {dx^{\mu_1}\over ds}\cdots
{dx^{\mu_{n}}\over ds}=T_{0\cdots 0}{dt\over ds}^{n}+{\rm kinetic~~ terms}
\ee
with $T$ a function of $F$, $R$ and their covariant derivatives.
 Near the horizon $({dt \over ds})^2 = G^{00}$ diverges. Nevertheless
$f$ must
remain regular,  or else  the scalar invariant
 $T_{\mu_1\ \cdots  \mu_{n}}T^{\mu_1\ \cdots  \mu_{n}}$  could not
possibly be  finite.

The only
remaining possibility is that
$f$ depends on extrinsic invariants,  or on other higher-derivative
terms
of the coordinate functions.
They are constructed by using the covariant accelerations $\Omega_n^{\mu}$.
For $n=1$, this is the usual four-velocity, $\Omega_1^{\mu}=\dot x^{\mu}$.
For $n=2$, this is the acceleration
\be
\Omega_2^{\mu}=\ddot x^{\mu}-{1\over 2}\partial_{\tau}\log(G_{\nu\rho}\dot
x^{\nu}\dot x^{\rho})\dot x^{\mu}
+\Gamma^{\mu}_{\nu\rho}\dot x^{\nu}\dot x^{\rho}
\ee
and so on.
Notice that $\Omega_n^{\mu}$ are tensors both of the target space diffeomorphisms
as well as world-line reparametrizations.
The piece of the acceleration that contributes to the potential is
\be
\Omega_2^{\mu}\left|_{pot}=-{1\over 2}{G_{00,0}\over G_{00}}\delta^{\mu}_{0}
+\Gamma^{\mu}_{00}\right.
\ee
This in principle can give singular contributions on the horizons via
invariants of the form $G_{\m\n}\Omega^{\m}_{m}\Omega^{\n}_{n}$.
We can show however that in the usual on-shell perturbative 
derivation of the higher 
$\alpha'$ corrections such invariants cannot appear.
The reason is that the first non-trivial correction is computed by 
matching $on$-$shell$ some scattering amplitude involving the particle
with a combination of higher velocity or acceleration terms.
However, on-shell $\Omega_{2}^{\m}\sim {F^{\m}}_{\n}\dot x^{\nu}$.
Thus, the $\Omega_2$ is redundant and does not appear on-shell in the first correction terms.
Consequently the corrected action involves only velocities, and the corrected equations will 
equate $\Omega_2$ to velocities again.
Thus, to any finite order of perturbation theory, we do not have the dangerous
acceleration terms.
This is supported by a recent calculation of (part of) the ${\cal O}(\alpha'^2)$ terms 
for D0-branes \cite{bbg}.

  The upshot  of the previous argument is that the right-hand
side of eq. (\ref{genrl}) is universal, and hence so is
the
differential equation (\ref{diff}) which one  can integrate for the
equation of state. One immediate corollary, assuming the
entropy stays smooth in the extremal limit where $T=0$, is that
$A_0(r_+) \Bigl|_{\rm extremal} = {1/l_p}$
always.

There is, to be sure,  still a lot of room for
string or quantum-gravity corrections to the thermodynamic functions.
  Both the chemical  potential,  $A_0(r_+)$, away from extremality,
and  the equation of state  for, say,  neutral holes   $S(M,0)$, are
expected in general to receive such corrections. It is also
conceivable that, like finite-size effects, string and/or  quantum
gravity corrections
invalidate our thermodynamic treatment of the problem.

\section{Near-extremal near-horizon limit}
\setcounter{equation}{0}

  To discard quantum gravity effects we will now assume that $l_p$ is
  vanishingly small compared to  all other length scales in the
  problem. We will furthermore take the near-extremal limit, and
  concentrate on the near-horizon geometry of the black hole,
\be{
l_p\;  \ll\;   \delta r \sim (r_+-r_-)\;  \ll\;  r_+ \ . \label{limit}
}
\ee
In order to analyze this limit, it
 is convenient to define the new coordinates
\be{
r \equiv  l_p^2 M \; (1+u)\ \ {\rm and}  \ \ t \equiv  l_p^2M \;
{\tilde t}\ ,
}
\ee
in terms of which  the metric  reads
\be
(l_p^2 M)^{-2}\; ds^2\; =\; -{f(u)\over \left(1+{1\over
  u}\right)^2}\; d{\tilde t}^2+
\left(1+{1\over u}\right)^2 \left({du^2\over
f(u)}+u^2 d\Omega_2^2\right) \ ,
\ee
where
\be
f(u) = 1 - ({u_0\over u})^2 \ \ {\rm and }\ \  r_{\pm} \equiv
  l_p^2M\; (1\pm u_0) \ .
\ee
The outer and inner horizons are  located in the new coordinates at $u
= \pm u_0$,
while   the electric potential takes the form
\be
A_0 \equiv (l_p^2 M)^{-1}\; {\tilde A}_0 = {\sqrt{1-u_0^2}\over l_p
  (1+u)}\ .
\ee
Finally the potential of a point probe can be worked out easily with
  the result
\be
 V(u) = m {\sqrt{u^2 - u_0^2}\over 1+u} + {q\over l_p}
  {\sqrt{1-u_0^2}\over 1+u}\ .
\ee

Consider now the limit
 (\ref{limit}) which can be written  equivalently as
$(l_p M)^{-1} \; \ll u, u_0 \; \ll \; 1$.
In this limit the metric simplifies to
 \be
(l_p^2 M)^{-2}\; ds^2\; =\; -{f(u) u^2 }
\; d{\tilde t}^2+
{du^2\over
f(u)u^2 }+  d\Omega_2^2 \ .
\ee
The extremal case ( $f=1$)  gives the $AdS_2\times S^2$ space,
also known  as the Bertotti-Robertson universe. For finite $u_0$, on
 the other hand,
one has a two-dimensional black hole embedded in this asymptotic
 geometry.\footnote{This solution is different from the one  discussed
in \cite{ads2}.}
For the thermodynamic quantities, we obtain\footnote{A similar result was
obtained in \cite{rn}.}
\be
S=\pi Q^2+4\pi^2Q^3 Tl_p+{\cal O}(l_p^2)
\label{en1}\ee
\be
U={Q\over l_p}+{Q^3\over 2}(2\pi T)^2~l_p+2Q^4(2\pi T)^3 l_p^2+{\cal
O}(l_p^3)
\ee
\be
\Phi={1\over l_p}-{2\pi TQ}-6\pi^2Q^2T^2~l_p+{\cal O}(l^2_p)
\ee
Note here that the leading contributions to $U$ and $\Phi$ are
singular.
This will also be the case for Dp-branes.

The static potential of a point probe reads in this
limit \cite{many}
\be
V(u) =  {q\over l_p} - {q\over l_p}u +  m \sqrt{u^2 - u_0^2}\ .
\ee
For a  generic probe $m> q/l_p$, so that the potential grows linearly
at the spatial infinity of $AdS_2$ space. For an extremal probe,  on
the other hand,
the potential goes to a constant at infinity, and the potential
difference $\delta V = V(\infty) - V(u_0)= {q\over l_p} u_0$  is well
defined. We can therefore use our thermodynamic argument to derive the
expression for  the
chemical potential
\be{
\mu = {1- u_0\over l_p}\ ,
}
\ee
 in agreement  with the
near-extremal expansion  of the electrostatic potential at the horizon
$\mu= A_0(u_0)$. Note that,  as with most  other
thermodynamic quantities, one has to keep the first subleading
correction  in the near-extremal  expansion of $\mu$,  in order to
find the leading temperature dependence.

  A heuristic rephrasing of the main message of this section  is as
follows: a
  near-extremal black hole exerts no net force on an
  extremal probe  at long distance. The  potential energy, is thus
  converted to kinetic energy and
  eventually released as heat while the probe falls in the
  near-horizon geometry. To leading order in the extremality parameter
  one can therefore compute the chemical potential by ignoring the
  physics in the asymptotically-flat  region.

\section{Black Dp-branes}
\setcounter{equation}{0}

We consider now the background geometry (in the string frame) of a
near-extremal black hole
describing a number of coinciding Dp-branes \cite{hs}:
\be
ds_{10}^2={-f(r)dt^2+d\vec x\cdot d\vec x\over
\sqrt{H_p(r)}}+\sqrt{H_p(r)}\left({dr^2\over
f(r)}+r^2d\Omega_{8-p}^2\right)\label{dmet}
\ee
where
\be
H_p(r)=1+{L^{7-p}\over r^{7-p}}\;\;\;,\;\;\;f(r)=1-{r_0^{7-p}\over
r^{7-p}}
\ee
The parameters $L$ and $r_0$ determine the AdS throat size
and the position of horizon, respectively. They are related
to the ADM mass $M$ and the (integer) Ramond-Ramond charge N
in the following way:
\be
M={\Omega_{8-p}V_p\over
2\kappa_{10}^2}~\left[(8-p)r_0^{7-p}+(7-p)L^{7-p}\right]~~,~~
N={(7-p)\Omega_{8-p}\over
2\kappa^2_{10}T_p}~L^{(7-p)/2}\sqrt{r_0^{7-p}+L^{7-p}},
\label{charge}\ee
where $\Omega_{n}$ is the volume of a unit $n$-dimensional sphere,
and $V_p$ is the common $p$-dimensional D-brane (flat) volume.
The  relations (\ref{charge})
involve the D-brane tension $T_p$ and the 10-dimensional
gravitational constant $\kappa_{10}$.
The RR charge N is quantized, with each D-brane carrying a unit charge
so that N is equal to the number of D-branes.
Finally,
\be
L^{7-p}=\sqrt{\left({2\kappa_{10}^2T_pN\over
(7-p)\Omega_{8-p}}\right)^2
+{1\over 4}r_0^{2(7-p)}}-{1\over 2}r_0^{7-p}
\ee

The RR charge is the source of the $p$-form field
\be
C_{012\cdots p}(r)~=~{2\kappa^2_{10}T_pN\over
\Omega_{8-p}(7-p)(r^{7-p}+L^{7-p})}~=~\sqrt{1+{r_0^{7-p}
\over L^{7-p}}}~{H_p(r)-1\over H_p(r)}\ .
\label{C}\ee
All other components vanish, except in the case of $p=3$,
when the self-duality condition
\be
{}F_{\mu_1\cdots\mu_5}={1\over
5!\sqrt{\det g}}
\epsilon_{\mu_1\cdots\mu_5\nu_1\cdots\nu_5}F^{\nu_1\cdots\nu_5}
\label{self}\ee
requires non-zero $p$-form components in the transverse
directions.
There is also a dilaton background (constant for $p=3$):
$e^{\phi}=H^{(3-p)/4}_p(r)
$.

By using standard methods of black hole thermodynamics,
it is straightforward to determine the
Hawking temperature,
entropy and chemical potential corresponding to the
solution (\ref{dmet},\ref{C}). They are respectively:
\be
T={7-p\over 4\pi}~{r_0^{(5-p)/2}\over \sqrt{r_0^{7-p}+L^{7-p}}}\
{}~~,\Phi=V_p T_p~{L^{(7-p)/2}\over \sqrt{r_0^{7-p}+L^{7-p}}}\label{th}
\ee
\be
S={4\pi\Omega_{8-p}V_p\over
2\kappa_{10}^2}~r_0^{(9-p)/2}\sqrt{r_0^{7-p}+L^{7-p}}\ ,
\ee

It is easy to check that these quantities satisfy the thermodynamic
identity$dU=TdS+\Phi dN$
with $U=M$.

We consider now a Dp-brane probing the above solution,
with zero background values for all other fields. In this case, the
D-brane
probe action is
\be
\Gamma_p=T_p~e^{-\phi}\int \sqrt{\det \hat g}+T_p\int \hat
C\label{gdp}
\ee
where we have also set the world-volume $F_{\alpha\beta}=0$.
Using the solution above
we obtain the static potential \cite{mal}
\be
V(r)=V_p T_p\left[{\sqrt{f(r)}\over H_p(r)}+C(r)\right]=V_p
T_p\left[{\sqrt{f(r)}\over H_p(r)}+\sqrt{1+{r_0^{7-p}\over
L^{7-p}}}{H_p(r)-1\over H_p(r)}\right]\label{4}\ee
where $C(r)\equiv C_{012\cdots p}(r)$. Note that, like in the RN case,
the horizon value $C(r_0)$ is equal to the chemical potential $\Phi$.
The values of the potential at infinity
and at the horizon are, respectively, $V(\infty)=V_pT_p$ and $V(r_0)=\Phi$.

A D$p$-brane probe is a BPS state with the mass $\Delta M=V_pT_p$
and charge $\Delta N=1$.
As it moves from infinity to the horizon,
its potential energy changes by $\Delta E=V(r_0)-V(\infty)$,
and the quantity of heat received by the black hole is again
$dE=-\Delta E$. On the other hand,
the black hole gains mass $dM=\Delta M=V_pT_p$ and charge $dN=\Delta
N=1$.
This process is described by the equation
\be
dE=V(\infty)-V(r_0)=dM-\Phi dN=dU-\Phi dN=T dS
\ee
which does indeed hold.
As expected, the probe motion, governed by the background field action
of eq.(\ref{gdp}), is consistent with black hole thermodynamics and provides
a similar partial equation of state as in the RN case.
The argument concerning the absence of $\alpha'$ corrections here is more involved.
Unlike the case of one-dimensional world-volumes, here the equations of 
motion do not set the accelerations (second fundamental forms) to be a 
function of velocities \cite{bbg}. 
At this point we can argue that  to order ${\cal O}(\alpha'^2)$ there are 
no corrections using the explicit
results of \cite{bbg}.
It turns out that the second fundamental form enters in such a way that there are
no extra corrections again on the horizon.
Moreover the appearance of a non-zero five form is not expected to change the 
previous statement.
Here we must also investigate the higher anomalous CP-odd couplings.
The relevant case is that of D3 branes and the coupling
\be
 S_{\theta}=-T_3{(4\pi\alpha')^2\over 48}\int a \left[p_1({\cal T})-
 p_1({\cal N})\right]
 \ee
where $p_1$ denotes the first Pontriagin class and ${\cal T},{\cal N}$
stand for
the tangent and normal bundles to the brane respectively
It can be shown that this vanishes for diagonal metrics with the required 
Poincar\'e symmetry.
Thus, there is no correction to the heat up to order ${\cal O}(\alpha'^2)$ 
and we suspect that this is true to all orders.
This would imply again that all corrections come from the corrected 
background fields.

\section{D3-branes in the near-horizon limit}
\setcounter{equation}{0}

The case of D3-branes is particularly interesting because
the world-volume action of $N$ coinciding D-branes
involves a four-dimensional ${\cal N}{=}4$
supersymmetric $SU(N)$ Yang-Mills theory. According to the Maldacena
conjecture
\cite{mald}, the large $N$ limit of this gauge theory is related
to the near-horizon AdS geometry of the extremal ($r_0=0$) black
D3-brane
solution (\ref{dmet}). Witten \cite{w2} has
exploited the AdS/SYM correspondence in order to study the large $N$
dynamics
of non-supersymmetric SYM, with ${\cal N}{=}4$ supersymmetries
broken by non-zero temperature effects. According to this proposal,
the non-extremal solution (\ref{dmet}) may be used to study
SYM at $T$ identified with the Hawking
temperature (\ref{th}) as long as $T\ll 1/L$, so that the metric
remains near-extremal $(r_0\ll L)$. In the Maldacena limit,
$\alpha'\equiv
l_s^2\to 0$ at $u\equiv r/\alpha'$ and $T$ fixed, the solution
(\ref{dmet})
describes an AdS-Schwarzschild black hole \cite{hp}:
\be
ds^2=l_s^2\bigg[{u^2\over R^2}(-f(u)dt^2+d\vec x\cdot d\vec x) +R^2
{du^2\over u^2f(u)}+R^2 d\Omega_{5}^2\bigg]+{\cal O}(l_s^4)\ ,\label{met}
\ee
where
\be f(u)=1-{u_0^{4}\over u^{4}}\qquad ,\qquad
R^4\equiv 4\pi g_s N=\lambda\qquad,\qquad u_0=\pi T R^2\, ,\ee
where $\lambda$ is the t'Hooft coupling.
The limiting value of the four-form (\ref{C}) is
\be
C_{0123}=1+l_s^4\left({(\pi T R)^4\over 2}-{u^4\over R^4}\right)+
{\cal O}(l_s^8),
\label{Cc}\ee
As in the near-horizon limit of the RN black-hole, here also the
gauge field diverges at the boundary of $AdS_5$, $u\to \infty$.

A D3-brane probe in the bulk of the AdS space corresponding
to $N$ background D3-branes can be thought of as a realization
of $SU(N+1)$ gauge theory in the $SU(N)\times U(1)$ symmetric Higgs
phase.
In the following, we examine some aspects of the  probe dynamics and
thermodynamics
in the near-horizon limit, in order to show that it is
well-defined and it leads to sensible results also in the Higgs phase.
To that end, we will use the following expansions in the string length
scale
$l_s$:
\be
L^4=R^4 l_s^4\left(1-{1\over 2}\pi^4R^4T^4l_s^4\right)+{\cal
O}(l_s^{12})~~,~~
r_0=\pi T R^2 l_s^2\left(1+{1\over 4}\pi^4 T^4 R^4l_s^4+{\cal
O}(l_s^{8})\right)\ .
\ee
which follow from relations written in the previous section.
Similarly,
\be
M=NV_3T_3+{3\over 8}\pi^2V_3N^2T^4+{\cal O}(l_s^4)~~,~~
S={1\over 2}\pi^2V_3 N^2 T^3+{\cal O}(l_s^4)\label{ent}
\ee
\be
\Phi=V_3T_3-{1\over 4}\pi^2V_3NT^4+{\cal O}(l_s^4)
\ee
Note that $T_3\sim 1/l_s^{4}\to \infty$. As pointed out before in
ref.\cite{gkt}, the limiting entropy (\ref{ent}) is 3/4 of the
corresponding quantity in the weakly coupled $SU(N)$ SYM.

Taking the limit in the static potential (\ref{4}), we
obtain\footnote{We
disagree with the potential obtained in the near-horizon limit in
\cite{ty}.}
\be
V(u)=V_3T_3\left\{1+l_s^4{u^4\over R^4}\left[\sqrt{1-\left({\pi T
R^2\over u}\right)^4}-1+{1\over 2}\left({\pi T R^2\over
u}\right)^4\right]+{\cal O}(l_s^8)\right\}
\ee
so that the interaction energy is
\be
V^{\rm int}(u)=V(u)-V_3T_3={V_3\over (2\pi)^3g_s}{u^4\over R^4}
\left[\sqrt{1-\left({\pi T
R^2\over u}\right)^4}-1+{1\over 2}\left({\pi T R^2\over
u}\right)^4\right]+{\cal O}(l_s^4)
\label{pot}\ee
and has a smooth limit as $l_s\to 0$.
Since the probe is BPS, $V^{\rm int}(\infty)=0$ as in the RN case.

The thermodynamic argument is still valid.
We let the probe fall until it reaches the horizon.
The heat supplied by the black hole to the probe is
\be
dE=V(\infty)-V(u_0)=V_3T_3-V(u_0)=\Delta M-\Phi \Delta N={1\over
4}\pi^2 V_3 NT^4
\ee
where $\Delta N=1$.
On the other hand, taking the limit of the equations of the previous
section we
obtain that the black hole cools down by $dT=-T/(2N)$
which is sufficient to prove directly that $dE=TdS$, with the entropy
given by eq.(\ref{ent}).
We thus find as before consistent thermodynamics, when we allow the probe
brane
to move from the boundary of AdS space to the horizon.
This (gravitational) equality corresponds to the qualitative 
expectation that an ultraviolet fluctuation in N=4 SYM  spreads until it reaches 
the size of the thermal wavelength and thus thermalizes \cite{bdhm}.

\newpage

\centerline{\large \bf Acknowledgements}
\vskip .6cm

We would like to thank C. Bachas for his collaboration 
and fruitful discussions.
E. Kiritsis would like to thank the Ecole Polytechnique
for hospitality and support while parts of this work was done.
He is also grateful to the SISSA node for financially
supporting his participation to the TMR midterm meeting.
The work of T.R. Taylor was done while he was visiting
Laboratoire de Physique Th\'eorique et Hautes Energies at Universit\'e
de Paris Sud, Orsay. He
is grateful to Pierre Bin\'etruy and all members of LPTHE
for their kind hospitality and support.

\def\O{\Omega}
\def\k{\Xi}
\def\m{\mu}
\def\n{\nu}


\begin{thebibliography}{99}

\bibitem{hs} G.T. Horowitz and A. Strominger, \np 360 (1991) 197.

\bibitem{gt} G.W. Gibbons and P.K. Townsend, \prl 71 (1993) 3754.

\bibitem{mald} J. Maldacena, hep-th/9711200.

\bibitem{gkp} S.S. Gubser, I.R. Klebanov and A.M. Polyakov,
hep-th/9802109.

\bibitem{w1} E. Witten, hep-th/9802150.

\bibitem{w2} E. Witten, hep-th/9803131.

\bibitem{hp} S.W. Hawking and D. Page, Comm.\ Math.\ Phys.\ 87 (1983)
577.

\bibitem{gkpeet} S.S. Gubser, I.R. Klebanov and A.W. Peet, \pr 54
(1996) 3915.

\bibitem{gkt} S.S. Gubser, I.R. Klebanov and A.A. Tseytlin,
hep-th/9805156.

\bibitem{bdhm} T. Banks, M. Douglas, G. Horowitz and E. Martinec,
hep-th/9808016\\
V. Balasubramanian, P. Kraus, A. Lawrence, S. Trivedi,  hep-th/9808017.

%\bibitem{pt} J. Pawelczyk and S. Theisen, hep-th/9808126.


\bibitem{bbg} C. P. Bachas, P. Bain, M. B. Green, hep-th/9903210.

\bibitem{bktt} C. P. Bachas, E. Kiritsis and T.R. Taylor, in preparation.
\bibitem{bhc} A. Strominger and C. Vafa,  hep-th/9601029, Phys. Lett.
{\bf B379} (1996) 99;\\
C. G. Callan and J. Maldacena, hep-th/9602043, Nucl. Phys. {\bf
B472} (1996) 591;\\
A. Dhar, G. Mandal and S. R. Wadia, hep-th/9605234, Phys.
Lett. {\bf B388} (1996) 51;\\
S. R. Das, and S. D. Mathur, hep-th/9606185, Nucl. Phys. {\bf B478}
(1996)
561; hep-th/9607149, Nucl. Phys. {\bf B482} (1996) 153;\\
J. Maldacena and A. Strominger, hep-th/9609026, Phys. Rev. {\bf D55}
(1997) 861; hep-th/9702015, Phys. Rev. {\bf D56} (1997) 4975;\\
S. F. Hassan and  S. R. Wadia, hep-th/9712213, Nucl. Phys. {\bf B526}
(1988) 311.

\bibitem{ads2} A. Strominger, hep-th/9809027.

\bibitem{rn} J. Maldacena, J. Michelson and A. Strominger,  JHEP 9902 (1999)
011; hep-th/9812073.
\bibitem{many} P. Claus, M. Derix, R. Kallosh, J. Kumar, P. Townsend
and
A. Van Proyen, hep-th/9804177.

\bibitem{mal} J. Maldacena, Phys. Rev. {\bf D57} (1998) 3736,
hep-th/9705053.



\bibitem{ty} A.A. Tseytlin and S. Yankielowicz, hep-th/9809032.


\end{thebibliography}
\end{document}